\documentclass[conference]{IEEEtran}
\IEEEoverridecommandlockouts

\usepackage{cite}
\usepackage{amsmath,amssymb,amsfonts}
\usepackage{algorithmic}
\usepackage{graphicx}
\usepackage{textcomp}
\usepackage{xcolor}
\usepackage{subcaption}
\usepackage{multirow}
\usepackage{array}
\usepackage{booktabs}
\usepackage{url}
\def\BibTeX{{\rm B\kern-.05em{\sc i\kern-.025em b}\kern-.08em
    T\kern-.1667em\lower.7ex\hbox{E}\kern-.125emX}}
\begin{document}








\title{Synecdoche: Efficient and Accurate In-Network Traffic Classification via Direct Packet Sequential Pattern Matching}

\author{\IEEEauthorblockN{Minyuan Xiao, Yunchun Li, Yuchen Zhao, Tong Guan, Mingyuan Xia, Wei Li*\thanks{*Wei Li is the Corresponding author.}}
\IEEEauthorblockN{Beihang University} 
\IEEEauthorblockN{\{xiaominyuan, lych, ruin1732,  guantong, catfish2339210,  liw\}@buaa.edu.cn}
}


\maketitle

\begin{abstract}
Traffic classification on programmable data plane holds great promise for line-rate processing, with methods evolving from per-packet to flow-level analysis for higher accuracy. However, a trade-off between accuracy and efficiency persists. Statistical feature-based methods align with hardware constraints but often exhibit limited accuracy, while online deep learning methods using packet sequential features achieve superior accuracy but require substantial computational resources. This paper presents Synecdoche, the first traffic classification framework that successfully deploys packet sequential features on a programmable data plane via pattern matching, achieving both high accuracy and efficiency. Our key insight is that discriminative information concentrates in short sub-sequences—termed Key Segments—that serve as compact traffic features for efficient data plane matching. Synecdoche employs an "offline discovery, online matching" paradigm: deep learning models automatically discover Key Segment patterns offline, which are then compiled into optimized table entries for direct data plane matching. Extensive experiments demonstrate Synecdoche's superior accuracy, improving F1-scores by up to 26.4\% against statistical methods and 18.3\% against online deep learning methods, while reducing latency by 13.0\% and achieving 79.2\% reduction in SRAM usage.
\end{abstract}

\begin{IEEEkeywords}
traffic classification, programmable data plane,  pattern matching.
\end{IEEEkeywords}

\section{Introduction}
The convergence of 5G, IoT, and edge computing is driving an evolution towards automated, intelligent networks. These networks demand real-time traffic analysis and classification capabilities for optimal performance \cite{b1,b2}. Such capabilities are critical for cybersecurity, enabling the rapid identification and mitigation of threats, as well as for guaranteeing the stringent Quality of Service (QoS) required by latency-sensitive applications. Consequently, the ability to perform accurate, real-time traffic classification directly within the data stream has become a cornerstone for the security, reliability, and performance of these next-generation networks.

To address these real-time requirements, performing traffic classification directly on the data plane offers significant advantages by avoiding the latency overhead of control plane round-trips and centralized processing bottlenecks. Programmable Data Planes (PDPs)\cite{b3, b4}, powered by languages like P4\cite{b5}, have emerged as a promising solution. They enable in-network inference by embedding classification models directly into the switching fabric, thus achieving high-throughput, line-rate traffic classification.

Traffic classification methods have evolved from per-packet approaches\cite{b9, b10, b11} to flow-based methods, as flow-level analysis provides richer contextual information and higher classification accuracy\cite{b7,b8}. Within flow-based approaches, existing methods can be categorized into two main paradigms based on the types of features they employ: statistical feature-based methods and packet sequential feature-based methods. Statistical feature-based approaches extract aggregated flow statistics and typically employ decision tree-based models\cite{b7, b8, b12}, which are compatible with programmable switches. However, these methods suffer from inherent accuracy limitations due to the limited expressiveness of statistical summaries\cite{b30,b31,b32,b33,b34}. In contrast, sequence feature-based approaches leverage raw packet content and ordering information, typically employing deep learning models to learn discriminative patterns \cite{b13,b14, b15}. However, deploying these models on the data plane presents significant challenges. Hardware constraints requiring model quantization can degrade performance\cite{b15}, while online inference consumes more hardware resources and introduces extra latency\cite{b6}.

This raises a compelling research question: Can we harness the high accuracy of sequential features while maintaining efficient classification on the data plane? Our key insight is that we can leverage offline analysis to discover discriminative sequential patterns. Traffic flows inherently contain critical sub-sequences—Key Segments\cite{b18, b19}—that encapsulate the essence of packet sequential features. These segments, typically spanning just a few consecutive packets, capture the distinctive communication patterns that differentiate various traffic classes. We can transform them into range-based match-action table entries, effectively distilling the discriminative power of entire packet sequences into compact, hardware-friendly matching rules.

To realize this insight, we propose Synecdoche, a novel traffic classification framework that bridges the accuracy-efficiency gap through an "\textbf{offline discovery, online matching}" paradigm. Synecdoche consists of two phases: First, \textbf{the discovery of Key Segments} (Section IV), where we employ deep learning models and Grad-CAM techniques to automatically extract discriminative packet sub-sequences from training data. Second, \textbf{matching with Key Segments} (Section V), where we transform discovered segments into optimized table entries and design a matching pipeline for line-rate classification. To the best of our knowledge, this represents the first work to successfully deploy sequential feature-based classification through direct pattern matching on a programmable data plane.

Our main contributions are:
\begin{itemize}
\item \textbf{Automated Key Segment Discovery Technique}: We propose a pipeline using deep learning and Grad-CAM that automatically discovers highly discriminative Key Segments from traffic data.
\item  \textbf{Efficient Data Plane Pattern Matching Framework}: We design the first efficient data plane architecture through direct pattern matching of packet sequential features, successfully achieving both high efficiency and high accuracy on a programmable data plane.
\item \textbf{Comprehensive System Implementation and Experimental Validation}: We implement Synecdoche on a Tofino switch and conduct extensive experiments on public traffic datasets, achieving F1-score improvements of up to 26.4\% over statistical methods and 18.3\% over online deep learning approaches, while reducing latency by 13.0\% and achieving 79.2\% reduction in SRAM usage compared to existing approaches.
\end{itemize}

\section{Background And Related Works}

\subsection{Programmable Data Plane}\label{AA}
The evolution towards more dynamic and intelligent networks has been significantly propelled by the advent of the programmable data plane. At the heart of this evolution is the Protocol-Independent Switch Architecture (PISA), which deviates from the rigid, fixed-function pipelines of traditional network switches. PISA provides the flexibility to define custom packet processing logic directly in hardware, enabling rapid deployment of new network services without costly hardware replacement cycles.

The PISA pipeline consists of three primary programmable blocks: 1) A Programmable Parser that identifies and extracts header fields from incoming packets according to user-defined protocol specifications; 2) A Match-Action Pipeline comprising multiple sequential stages, each equipped with memory (SRAM, TCAM) and ALUs, where packets are matched against flow tables and subjected to corresponding actions; 3) A Programmable Deparser that reconstructs packets from processed headers and payload before transmission. This architectural flexibility is programmable via domain-specific languages like P4, offering immense opportunities. However, it also imposes strict constraints to maintain line-rate processing.

Any practical data plane solution must operate within these limitations:

\textbf{Memory Limitation}: High-speed memory in each stage is scarce. TCAM for complex matches is particularly limited, while SRAM for exact-match tables is more abundant but still orders of magnitude smaller than conventional server memory.

\textbf{Operation Restrictions}: To ensure microsecond-level latency, ALUs support only minimalistic operations. Complex instructions such as multiplication, division, and floating-point arithmetic are unsupported, requiring decomposition into simple integer and bitwise operations.

These constraints collectively make direct deployment of conventional machine learning models a formidable challenge, necessitating novel approaches specifically designed for this unique computational environment.

\subsection{Related Work}
Existing traffic classification approaches on programmable data plane can be categorized into two primary paradigms based on their processing methodology: \textbf{per-packet methods} and \textbf{flow-level methods}. Per-packet methods perform classification directly on individual packet features without storing stateful information. For instance, Leo \cite{b9} proposed a sub-tree multiplexing mechanism to implement scalable and runtime-programmable online decision tree models. Mousika \cite{b10} utilized knowledge distillation to convert complex ML models into hardware-friendly binary decision trees. However, due to the limited information available in individual packets, per-packet methods often exhibit limited accuracy\cite{b7,b8}. To address this limitation, flow-level methods have emerged, which can be further divided into two subcategories.

\textbf{Statistical Feature-Based Methods}: These approaches on programmable data plane extract aggregated flow statistics and employ decision tree-based models, as their hierarchical rule structure can be directly translated into the match-action tables native to programmable switches. Flowrest \cite{b8} fully implemented a random forest model on switches to support more complex flow-level inference.NetBeacon \cite{b7} proposed a multi-phase sequential model architecture that dynamically analyzes packets with line-speed flow-level features while addressing deployment scalability through efficient model representation and stateful storage management. SentinelX \cite{b12} introduced TreeDivider and DualTree algorithms that achieve significant space reduction and improved detection accuracy through optimized flow table representation and dual-threshold decision trees.

\textbf{Packet Sequential Feature-Based Methods}: These approaches analyze network traffic by extracting features from raw packet content and using temporal ordering information to capture sequential patterns and dependencies between packets. Such methods typically employ deep learning models to process feature sequences \cite{b25,b26,b27,b28,b29}. Existing work on programmable switches also deploys deep learning models to the data plane for sequential pattern learning. Brain-on-Switch \cite{b15} designed a data-plane-friendly RNN architecture enabling line-rate neural network-driven traffic analysis. RIDS \cite{b13} explored the co-design of RNN and programmable switches to build advanced intrusion detection systems. Quark \cite{b14} leveraged model pruning and quantization to fully offload CNN inference to the data plane. Linc\cite{b39} design a divide-and-conquer strategy for incremental model updates on data plane.

\subsection{Discussion}
Current traffic classification approaches face an accuracy-efficiency trade-off on the programmable data plane. Statistical feature-based methods suffer from two fundamental limitations: first, their performance is heavily reliant on handcrafted features, which restricts their generalizability\cite{b30,b31}; second, the aggregation of raw data into statistical features causes an inevitable loss of fine-grained information, which ultimately limits their classification accuracy\cite{b32,b33,b34}. In contrast, packet sequential feature-based approaches face significant deployment challenges on the programmable data plane. Hardware constraints necessitate model simplification strategies such as pruning and quantization, which degrade model classification accuracy \cite{b6}. Furthermore, deploying these models directly on the data plane consumes substantial memory and computational resources, introducing extra processing latency\cite{b15}. Therefore, efficiently leveraging packet sequential features on the data plane remains a challenge.

\hspace*{\fill}

\section{Key Segments: Definition, Properties, and Framework Overview}
This section defines our core concept of "Key Segments" and presents an overview of our framework for their discovery and deployment on a programmable data plane.

\subsection{Key Segments: Definition and Core Properties}
\begin{figure}[!htb]
\centering
\begin{subfigure}[b]{0.22\textwidth}  
    \centering
    \includegraphics[width=\textwidth]{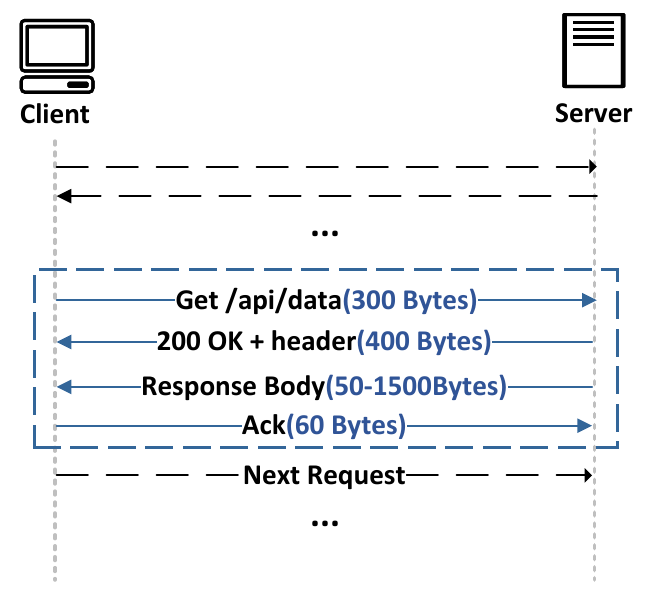}
    \caption{HTTPS}
\end{subfigure}
\quad  
\begin{subfigure}[b]{0.22\textwidth}
    \centering
    \includegraphics[width=\textwidth]{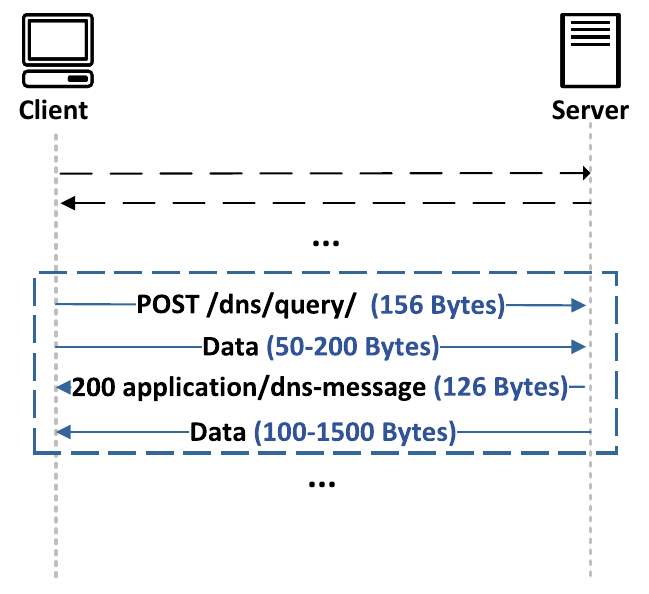}
    \caption{DoH}
\end{subfigure}
\caption{Examples of Key Segments}
\label{fig_ks}
\end{figure}

In traffic flows, specific packet patterns emerge due to the inherent structure of network protocols and application behaviors. These patterns, which we term \textbf{Key Segments}, originate from fundamental network operations such as protocol handshakes, request-response cycles, and application-specific data exchanges. As illustrated in Figure~\ref{fig_ks}, different applications exhibit distinct communication patterns. For example, HTTPS may show characteristic GET request-response sequences, while DoH can demonstrate compact POST-based DNS query patterns with predictable response structures.

Formally, given a traffic sequence $X = (x_1, \ldots, x_n)$, where each $x_i$ represents packet features, a Key Segment is a contiguous subsequence that makes significant contributions to identifying $X$ as belonging to a specific class. These segments capture the quintessential interaction patterns that distinguish different applications in traffic. The compactness of Key Segments makes them particularly suitable for hardware-based pattern matching implementations.

Key Segments are characterized by three core properties that make them both valuable for classification and challenging to discover:
\begin{enumerate}
    \item \textbf{Variable Length:} Segments range from short 2-packet sequences to complex multi-packet sequences, reflecting diverse application logic.
    \item \textbf{Positional Flexibility:} Key Segments appear at varying offsets within flows due to packet retransmissions and user behavior.
    \item \textbf{Value Variability:} Packet sizes within segments fluctuate based on factors like content length, though some packets (e.g., ACKs) maintain stable sizes.
\end{enumerate}

While these properties provide Key Segments with the expressive capability needed to effectively distinguish between different categories of traffic, they simultaneously create significant computational challenges for segment identification. The combinatorial search space across variable lengths, positions, and values leads to exponential complexity. Even with various restrictions such as limiting maximum segment lengths, the time complexity remains prohibitively high ~\cite{b18}.
\begin{figure*}[!t]
\centering
\includegraphics[width=\textwidth]{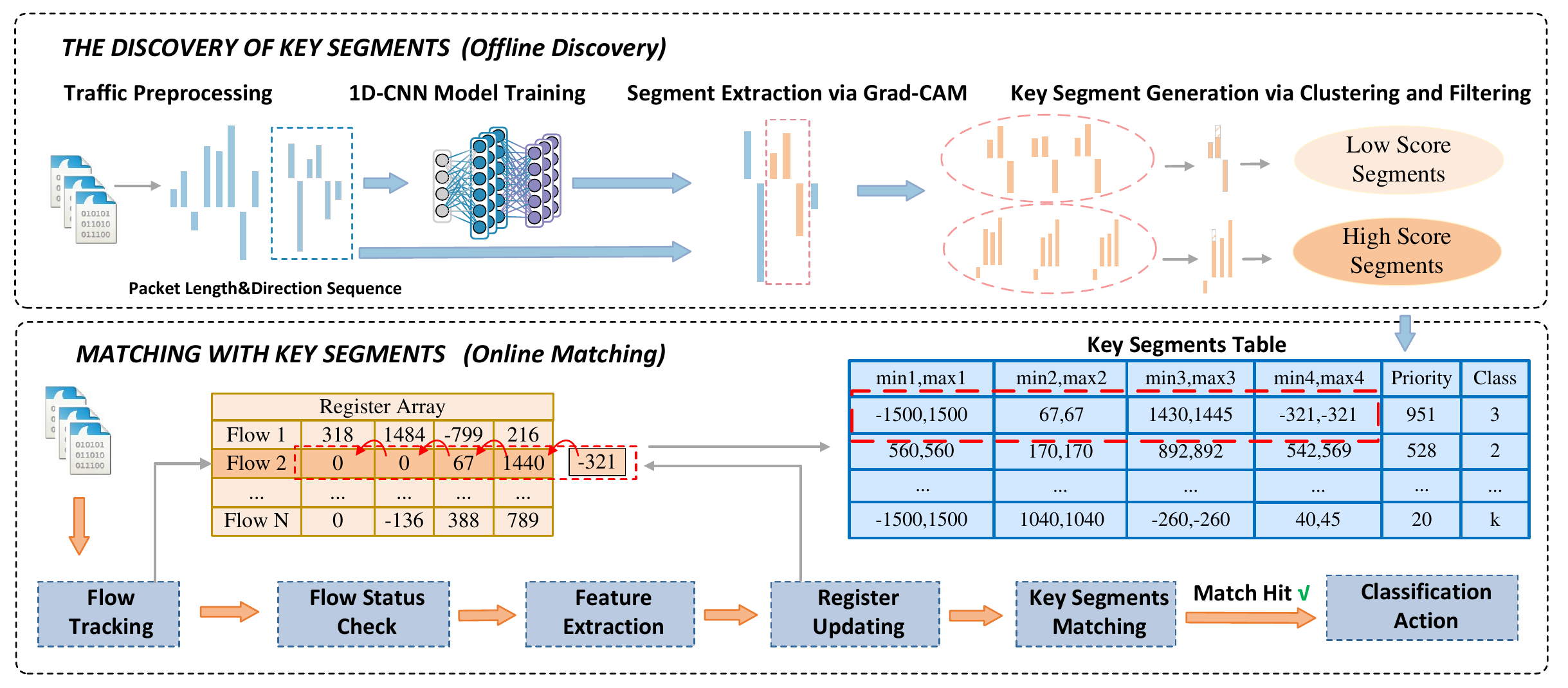}
\caption{The high-level architecture of our Synecdoche framework, illustrating the two-phase pipeline from offline discovery to online matching on the programmable data plane.}
\label{fig_framework}
\end{figure*}
\subsection{Framework Overview}
Figure~\ref{fig_framework} illustrates the overall architecture of our Synecdoche framework.

In the first phase, \textit{The discovery of Key Segments}, we tackle the challenge of automatically and efficiently identifying Key Segments from raw traffic. For this, we leverage a 1D-CNN model and Grad-CAM\cite{b20} techniques to pinpoint the most influential traffic segments for each class. The segments are then refined through clustering and filtering to generate a representative set of Key Segments.

In the second phase, \textit{matching with Key Segments on the data plane}, our framework transforms the discovered Key Segments into multi-key table entries and deploys them onto a programmable switch as a Key Segments Table. During online operation, the switch continuously updates packet feature registers for each flow and performs matching against the Key Segments Table. Once a successful match is detected, the flow is immediately classified, enabling dynamic and low-latency decision-making. Additionally, we provide a backup mechanism using a trained decision tree model to handle flows that fail to match any Key Segment within a predefined time window.

\section{The Discovery of Key Segments}

In this section, we elaborate on how to discover Key Segments from raw traffic.

\subsection{Traffic Preprocessing}
The process begins with preprocessing raw network data. We partition traffic into bidirectional flows based on the five-tuple (source/destination IP, source/destination port, protocol), then we extract the sequence of packet sizes with directional information. This feature representation has been proven effective for traffic analysis~\cite{b18,b19,b25,b26}. Specifically, for each bidirectional flow, we define $l_i$ as the length of the $i$-th packet, and $d_i$ as the direction using $\{+1, -1\}$ to represent the two directions. The combined feature is computed as:
\begin{equation}
x_i = l_i \times d_i
\end{equation}
Each flow is then represented as $\mathbf{X} = \{x_1, \ldots, x_k\}$, where $k$ is the number of packets in the sample flow.

\subsection{1D-CNN Model Training}
To learn the discriminative patterns within these sequences, we designed a 1D-CNN model for traffic classification. Its architecture consists of three parts: an embedding layer, a 1D-CNN backbone, and a classifier. The embedding layer treats each packet length as a discrete categorical variable rather than a continuous numerical value, converting the signed integers into dense vectors through embedding layers. This separation of classification meaning from numerical value makes the neural network easier to train and captures richer semantic relationships between different packet sizes. The core of our model is its 1D-CNN backbone, which excels at identifying local, position-variant patterns, a perfect analogue for our Key Segments. The final classifier uses softmax loss for the multi-class classification task. We train this 1D-CNN model following standard procedures for multi-class classification tasks.

\subsection{Segment Extraction via Grad-CAM}
With an accurately trained 1D-CNN model, we treat it as a learned knowledge base and apply model interpretability algorithms to identify which parts of an input sequence were most influential in its classification. We employ Grad-CAM~\cite{b20}, a gradient-based interpretability method that computes the gradient of the class score with respect to feature maps of the target convolutional layer, then uses global average pooling to obtain importance weights for generating class-discriminative localization maps.

\begin{figure}[!t]
\centering
\includegraphics[width=3.5in]{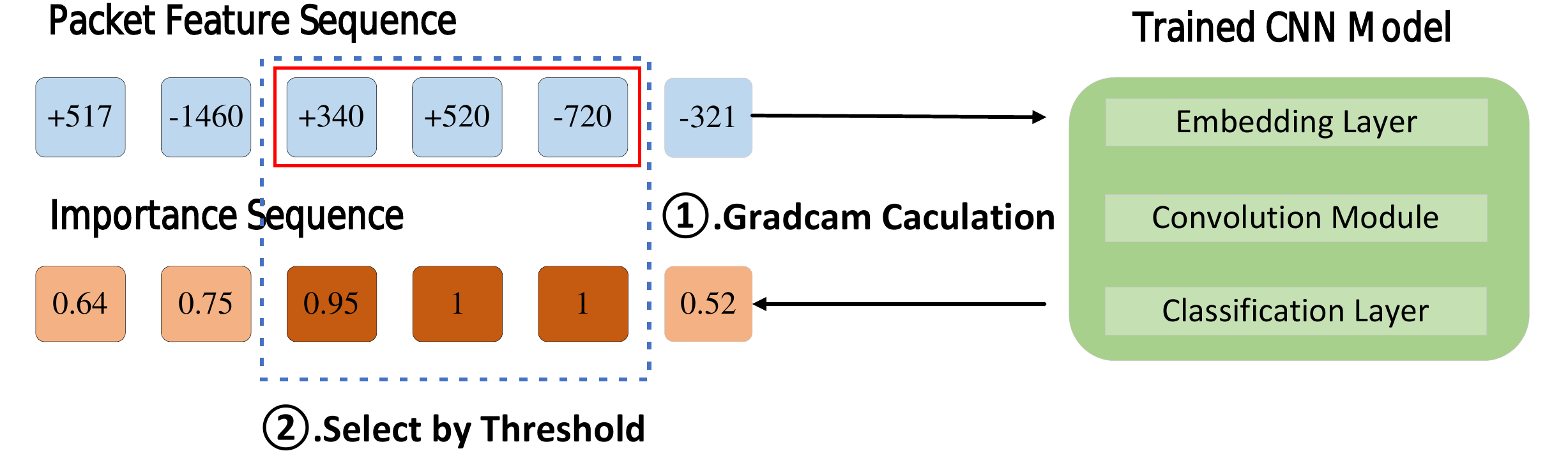}
\caption{An example of segment Extraction via Grad-CAM}
\label{fig_properties}
\end{figure}

The extraction process begins by applying Grad-CAM to compute an interpretability map for samples from each traffic category, assigning a continuous importance score to each packet position in the original input sequence. As illustrated in Figure~\ref{fig_properties}, given a packet feature sequence, Grad-CAM generates corresponding importance scores that highlight the most discriminative patterns. For the aggregated importance scores $S = \{s_1, s_2, \ldots, s_n\}$, we identify contiguous subsequences $\{x_i, x_{i+1}, \ldots, x_j\}$ from the original feature sequence where all corresponding importance scores satisfy:
\begin{equation}
s_k > \mu + t \cdot \sigma, \quad \forall k \in [i, j]
\end{equation}
where $\mu$ is the mean of all importance scores, $\sigma$ is the standard deviation, and $t$ is a threshold parameter.

We apply this extraction procedure to every sample in the training set to build a comprehensive pool of candidate segments. For each training sample, after identifying all contiguous subsequences that meet the threshold criterion, we apply length-based processing: subsequences shorter than the predefined minimum length $L_{min}$ are discarded as insufficiently discriminative. For subsequences within the acceptable range [$L_{min}$, $L_{max}$], we retain them entirely as candidate segments. When a subsequence exceeds the maximum length $L_{max}$, we use a sliding window to extract the sub-subsequence of length $L_{max}$ that achieves the highest cumulative importance score within it.

\subsection{Key Segment Generation via Clustering and Filtering}
The raw candidate segments, while informative, can be numerous and contain statistical noise. To generate representative Key Segments from the raw candidate segments, we employ a two-stage generation process.

First, candidate segments are grouped by their originating sample class. Within each class-based group, we pad all segments to the maximum length $L_{max}$ to capture segments of variable lengths as mentioned in our Key Segment properties. We then apply the DBSCAN clustering algorithm to identify dense regions of similar segments within each class. For each resulting cluster, we generate a single, generalized Key Segment as a \textbf{range-based template} directly suitable for data plane matching. For each position $i$ within the segments of a cluster, we compute the minimum and maximum observed packet feature values:

\begin{equation}
\text{Key Segment} = [(x_1^{min}, x_1^{max}), \ldots, (x_{L_{max}}^{min}, x_{L_{max}}^{max})]
\end{equation}

where $x_i^{min}$ and $x_i^{max}$ represent the minimum and maximum feature values at position $i$ across all segments in the cluster.

Second, since some of the segments might match multiple classes and lack sufficient discriminative power, we introduce a scoring mechanism to select only the most representative segments for each class. For each generated segment, we evaluate its discriminative power using a held-out validation dataset containing samples from all classes. For each segment $s$ and its target class, we calculate $C_{in}(s)$ as the fraction of in-class samples that match segment $s$. We also compute $C_{out}(c, s)$ as the fraction of samples in each other class $c$ that match segment $s$, and define:

\begin{equation}
C_{out}^*(s) = \max_{c \neq \text{target class}} C_{out}(c, s)
\end{equation}

The discriminative score for segment $s$ is then computed as:

\begin{equation}
\text{Score}(s) = \frac{C_{in}(s)}{C_{out}^*(s) + \epsilon}
\end{equation}

where $\epsilon$ is a small constant to prevent division by zero. This ratio-based scoring mechanism favors segments that achieve high coverage within their target class while maintaining low coverage across other classes. Only segments with scores exceeding a predefined threshold $S$ are retained for deployment. This ensures that only high-fidelity Key Segments with strong discriminative power are passed to the deployment stage.

\section{Matching with Key Segments}
In this section, we explain how to deploy discovered Key Segments on the data plane and how to use them in online classification.

\subsection{Key Segment Table Generation}
The first step in deployment is to translate the Key Segments generated by our offline discovery process into match-action table entries consumable by the P4 data plane.

Each discovered Key Segment is compiled into a single P4 table entry, with each position within the segment becoming a key field that leverages the TCAM's range matching capabilities on P4-programmable switches. Specifically, for a Key Segment $[(x_1^{min}, x_1^{max}), (x_2^{min}, x_2^{max}), \ldots, (x_{L_{max}}^{min}, x_{L_{max}}^{max})]$, we create $L_{max}$ key fields. Each key field $i$ is set to the range $[x_i^{min}, x_i^{max}]$ for meaningful positions, while positions that were padded during clustering are set to match the universal set, effectively creating wildcard matches. The priority of each table entry is set to the discriminative Score computed during the filtering phase, ensuring that more discriminative segments are matched first. The action associated with each entry writes the corresponding class ID of the Key Segment into the packet's metadata.

Additionally, we provide an SRAM version that decomposes each range-containing segment into multiple exact-match entries using Cartesian product expansion, enabling precise matching on hardware with limited range matching support. In the SRAM version, since universal set matching is not feasible, we create separate tables for each possible segment length.

\subsection{Packet Processing Pipeline}
The core of our system is the P4 packet processing pipeline that performs stateful matching at line rate. This pipeline design enables dynamic, early classification. A flow can be identified as soon as a matching Key Segment appears. The journey of each packet is as follows:

1) \textbf{Flow Identification}: Upon ingress, the packet is parsed to extract its 5-tuple. To handle bidirectional flows, we apply symmetric hashing to the 5-tuple, ensuring that packets from both directions of the same connection map to the same flow ID. This flow ID serves as an index for all subsequent stateful operations.

2) \textbf{Flow Status Check}: We query a flow classification table with its 5-tuple to determine whether this flow has already been classified. If the flow is found in the classification table with an assigned class label, the remaining steps are bypassed to conserve switch resources.

3) \textbf{Feature Extraction}: For each packet, we extract both length and direction information. The packet length is obtained directly, while the direction is determined by comparing the current packet's source IP with the first packet's source IP in the flow. The combined feature of length and direction is computed using the same addition-based approach as described in the preprocessing phase.

\begin{figure}[!t]
\centering
\includegraphics[width=3.5in]{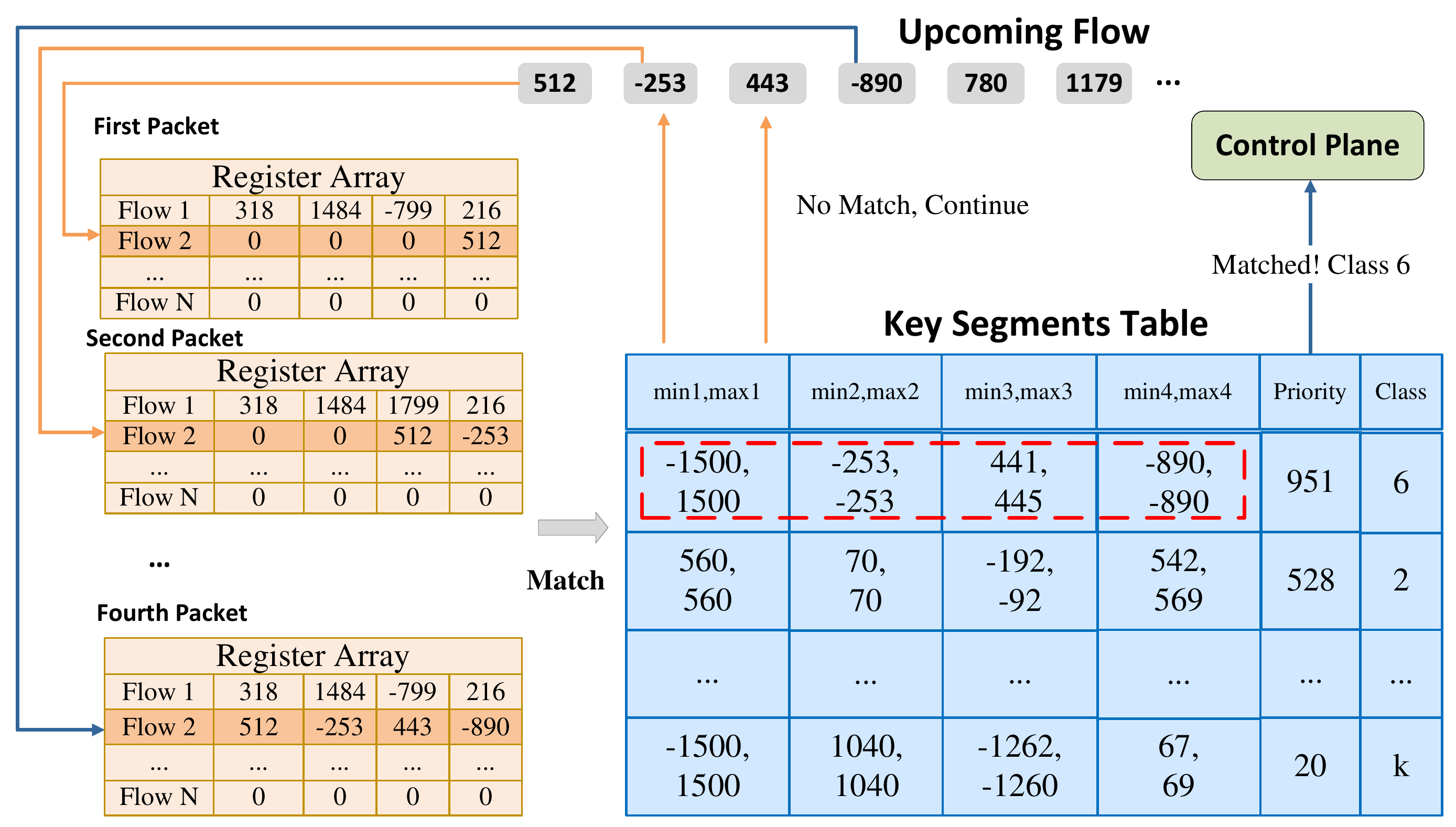}
\caption{An example of register update and Key Segment table matching}
\label{matching}
\end{figure}

4) \textbf{Stateful Register Update}: We maintain a large array of stateful registers on the switch, where each flow is mapped to its own dedicated register block of size $L_{max}$ via its flow ID. This register block functions as a sliding window, holding the feature values of the last $L_{max}$ packets observed for that flow. When a new packet arrives, the contents of the corresponding register block are shifted one position to the left, discarding the oldest packet feature, and the combined feature of the current packet is inserted into the newest position.As illustrated in Figure~\ref{matching}, flow 2's register array starts as $[0, 0, 0, 0]$, evolves to $[0, 0, 0, 512]$ after the first packet, then to $[0, 0, 512, -253]$ after the second packet, and so on.

5) \textbf{Key Segment Matching}: After the register is updated, its entire $L_{max}$-element content is used as the lookup key for the main Key Segment Table. If no match occurs, the process repeats when the next packet in the flow arrives.  As shown in Figure~\ref{matching}, for example, the first three packets of flow 2 result in register contents that do not match any table entries. However, when the fourth packet arrives and the register content becomes $[512, -253, 443, -890]$, it matches the first rule of the Key Segments Table, triggering classification and assigning the flow to Class 6.

6) \textbf{Classification Action}: If the lookup yields a match, the system uploads summary information to the control plane, including the flow's 5-tuple and the identified class. The control plane then installs the corresponding flow classification entry into the data plane's flow classification table. 

\subsection{Backup Classification Mechanism}
To ensure that all flows are eventually classified and to handle traffic that may not contain any of the discovered Key Segments, we incorporate a robust backup mechanism. The backup mechanism employs a pre-trained Decision Tree (DT) implemented on the data plane. This DT is trained offline using the same packet features (the first $L_{max}$ packets' length and direction) as Key Segments matching, requiring no additional storage space. 

The decision tree classification is performed when the $L_{max}$ packet arrives. The classification result is then stored and applied as the final flow label under two triggering conditions: (1) when the flow's packet count exceeds a predefined threshold $pkt_{max}$, or (2) when the time interval between consecutive packets in a flow exceeds $time_{max}$. In both cases, if no Key Segment has been matched yet, the system retrieves and uses the stored decision tree classification result as the final flow label.

\section{Evaluation}

In this section, we present a comprehensive evaluation of Synecdoche. Specifically, we aim to answer the following key research questions:

\textbf{(RQ1) Accuracy}: Does Synecdoche's direct matching of Key Segments achieve classification accuracy comparable to sophisticated online deep learning methods while surpassing traditional statistical feature-based approaches?

\textbf{(RQ2) Efficiency}: Can Synecdoche deliver superior performance in terms of latency and hardware resource utilization compared to existing data plane inference methods?

\subsection{Experimental Setup}

\textbf{Environment.}Our experiments are conducted in a two-part environment reflecting our framework's offline and online phases.
\begin{itemize}
    \item \textbf{Offline Training Environment:} The discovery of Key Segments, including the training of the CNN model and the subsequent refinement process, is performed on a server running Linux Ubuntu 22.04. The server is equipped with an Intel(R) Xeon(R) Gold 6330 CPU @ 2.00GHz and an NVIDIA RTX 4090 GPU. The software stack consists of Python 3.8, TensorFlow 2.6 for the deep learning framework, and the nfstream\cite{b35} for traffic processing. 
    \item \textbf{Online Data Plane Environment:} The line-rate classification engine is deployed and evaluated on a Ruijie RG-F9500-32CQ switch, which is equipped with an Intel Tofino ASIC. The data plane logic is implemented in P4\textsubscript{16}.
\end{itemize}

\textbf{Datasets.}We evaluate our method on the following tasks:
\begin{itemize}
    \item \textbf{IoT Cybersecurity}: We use two datasets to evaluate attack detection capabilities on IoT devices: (1)\textbf{Bot-IoT}\cite{b21} dataset containing normal and attack traffic from IoT devices with 4 classes (DoS, DDoS, Scan, Theft), and (2) \textbf{ToN-IoT}\cite{b22} dataset providing heterogeneous IoT network attack scenarios with 10 classes covering normal traffic and various attacks.
    \item \textbf{Application Classification}: We evaluate on two encrypted traffic datasets with varying class complexities: (1) \textbf{CipherSpectrum}\cite{b23} dataset includes network traffic encrypted with modern TLS 1.3 cipher suites for application domains, which we evaluate at four different scales: 5, 10, 20, and 42 (all)  website classes to analyze scalability. (2) \textbf{VisQUIC}\cite{b24} dataset containing QUIC-encrypted traffic from 16 different websites, which we evaluate at three scales: 5, 10, and 16 (all) website classes. CipherSpectrum-10 and VisQUIC-10 refer to the 10-class versions of each dataset.

\end{itemize}

For each dataset, we use a standard 80\%/10\%/10\% split for training, validating, and testing, respectively.

\textbf{Comparison Methods.}To comprehensively evaluate our approach, we compare Synecdoche against both data plane methods and offline baselines:
\begin{itemize}
\item Data Plane Methods: \textbf{NetBeacon}~\cite{b7}: State-of-the-art statistical feature-based approach using decision trees on programmable switches, representing traditional ML-based P4 inference. \textbf{Brain-on-Switch}~\cite{b15}: A representative framework for deploying quantized neural networks onto the data plane. 

\item Offline Baselines (Upper Bound Analysis): \textbf{Random Forest}: Offline statistical method using the same features as NetBeacon but without hardware constraints (unlimited tree depth and estimators). \textbf{FS-Net}~\cite{b31}: State-of-the-art offline sequence feature-based deep learning method, representing the upper bound of accuracy for sequence-based approaches.
\end{itemize}

To ensure fair comparison, we configure the input features and hyperparameters of comparing methods according to the settings specified in their respective official implementations\cite{b37,b38, b36}. For the IMIS system in BoS, we evaluate only the data plane's classification capability, excluding gains from data-control plane interaction, since Synecdoche can likewise be integrated into such systems.

\textbf{Evaluation Metrics.}We assess performance using two categories of metrics:

\begin{itemize}
    \item \textbf{Accuracy Metrics:} We use standard classification metrics, including overall Accuracy (Acc.) and the macro-averaged F1-Score (F1), which provides a balanced measure of precision and recall.
    
    \item \textbf{Performance Metrics:} Per-packet Latency (ns), and critical on-chip resource consumption (SRAM and TCAM as percentage of total available resources on Tofino ASIC).
\end{itemize}

\subsection{Hyperparameter Settings and Method Analysis}
In this subsection, we provide an in-depth analysis of our method's key components and validate the effectiveness of our Key Segment discovery approach through systematic parameter analysis and early classification validation.

Our framework involves several key parameters that affect both Key Segment matching accuracy and data plane efficiency. Through systematic parameter tuning, we maintain consistent parameter settings across most datasets. Table~\ref{tab:hyperparams} summarizes these key parameters. However, the score threshold $S$ requires dataset-specific optimization, as it controls the critical trade-off between first-stage matching rate and overall accuracy.

\begin{table}[h]
\centering
\caption{Key Hyperparameter Settings for Synecdoche}
\label{tab:hyperparams}
\renewcommand{\arraystretch}{1}
\begin{tabular}{@{}cc@{}}
\toprule
\textbf{Parameter}  & \textbf{Optimal Value} \\
\midrule
Min/Max Segment Length ($L_{\min},L_{\max}$) & 2, 4 \\
CNN Embedding Dimension & 128 \\
CNN Kernel Size / Feature Channel & 3, 128\\ 
Extraction Threshold ($t$) & 0.5 \\
Backup Mechanism Threshold ($\text{pkt}_{\max},\text{time}_{\max}$) & 30, 256ms \\
\bottomrule
\end{tabular}
\end{table}

\begin{figure}[!htb]
\centering
\begin{subfigure}[b]{0.22\textwidth}  
    \centering
    \includegraphics[width=\textwidth]{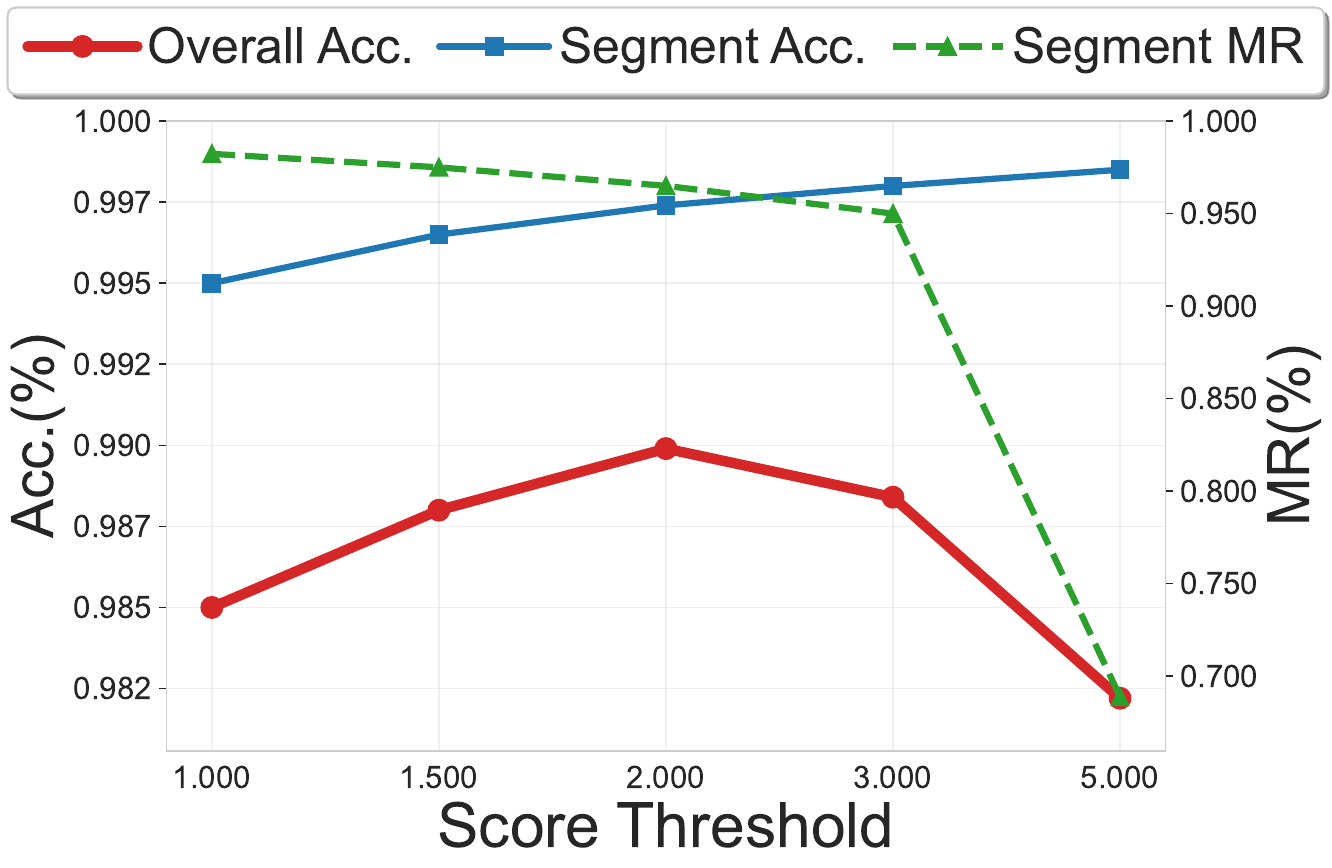}
    \caption{Bot-IoT}
\end{subfigure}
\quad  
\begin{subfigure}[b]{0.22\textwidth}
    \centering
    \includegraphics[width=\textwidth]{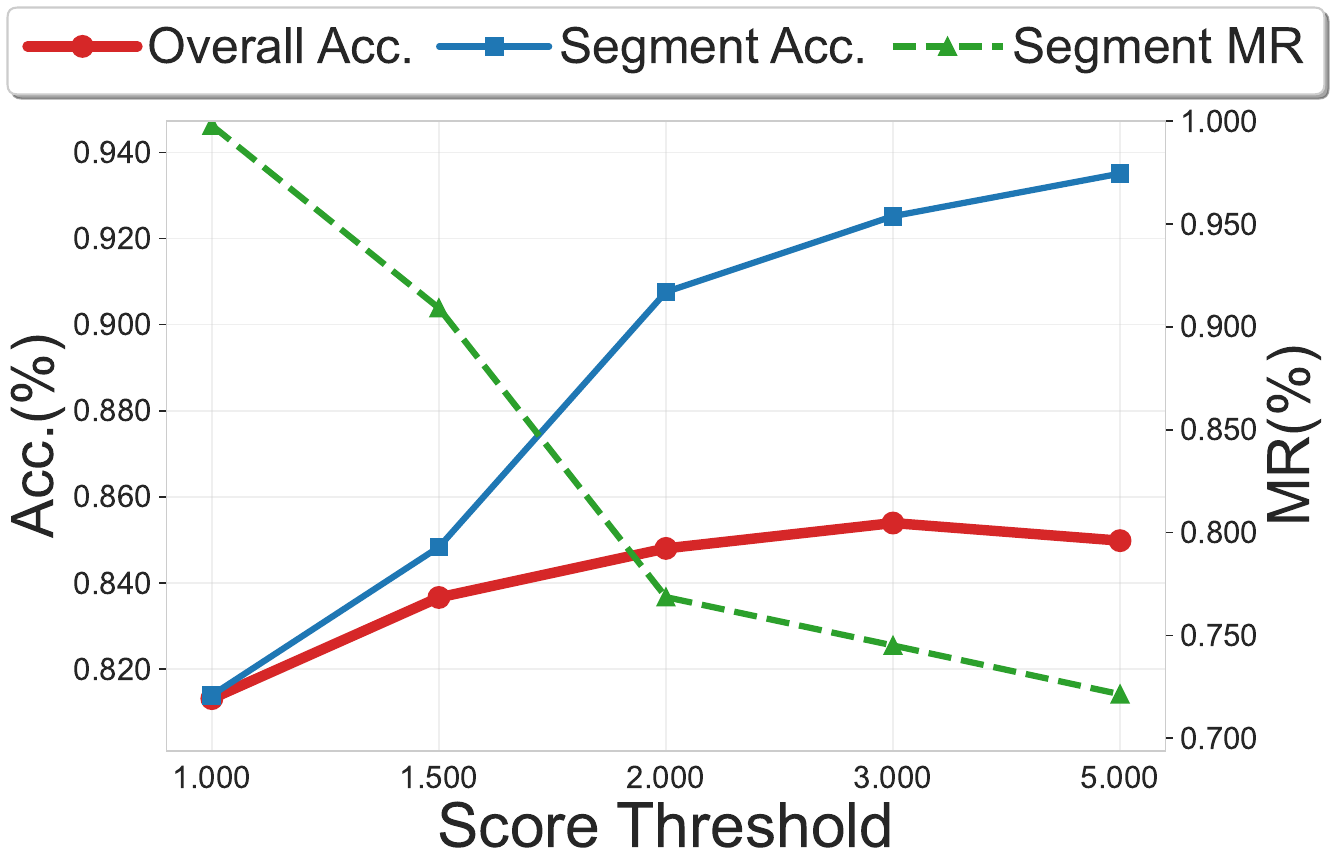}
    \caption{ToN-IoT}
\end{subfigure}

\medskip  

\begin{subfigure}[b]{0.22\textwidth}
    \centering
    \includegraphics[width=\textwidth]{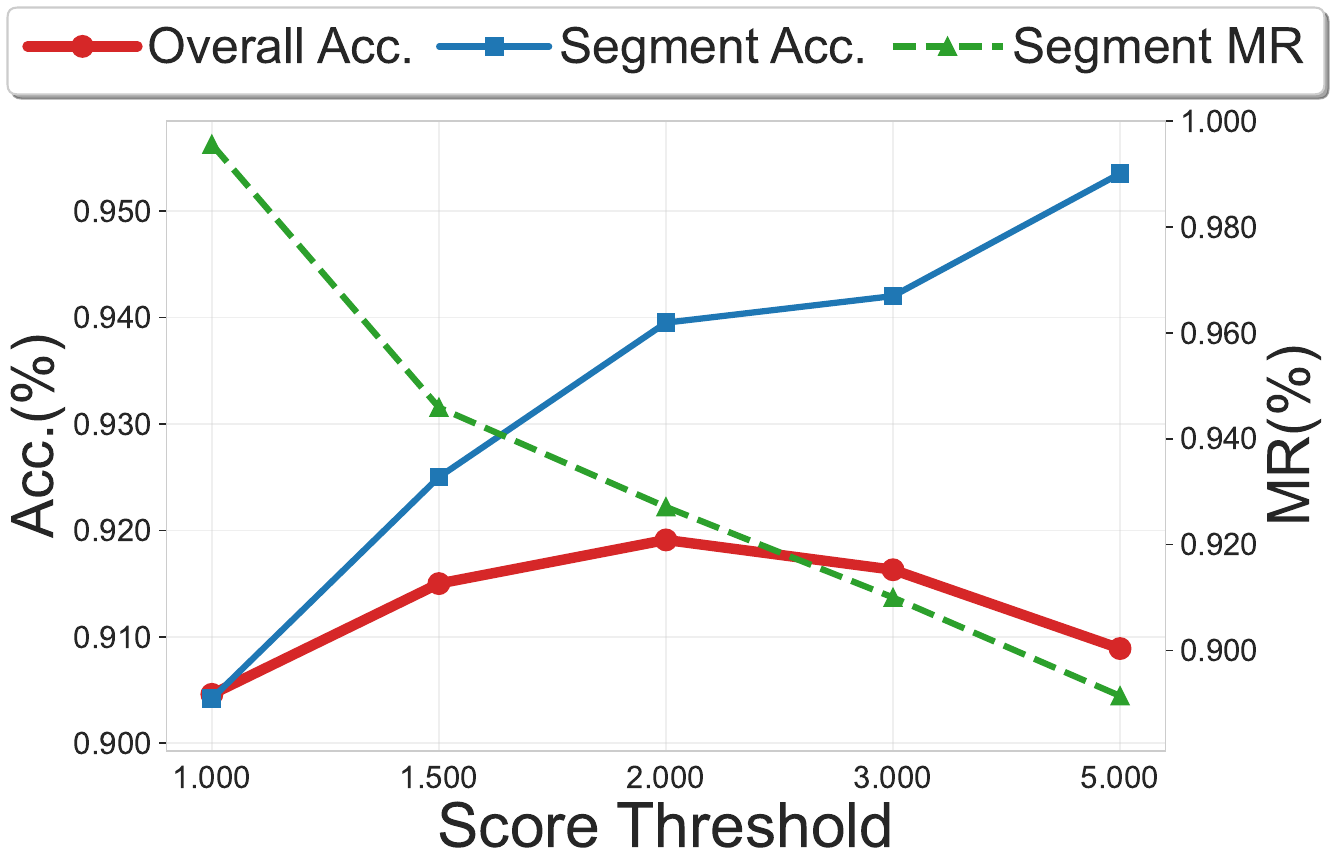}
    \caption{CipherSpectrum-10}
\end{subfigure}
\quad
\begin{subfigure}[b]{0.22\textwidth}
    \centering
    \includegraphics[width=\textwidth]{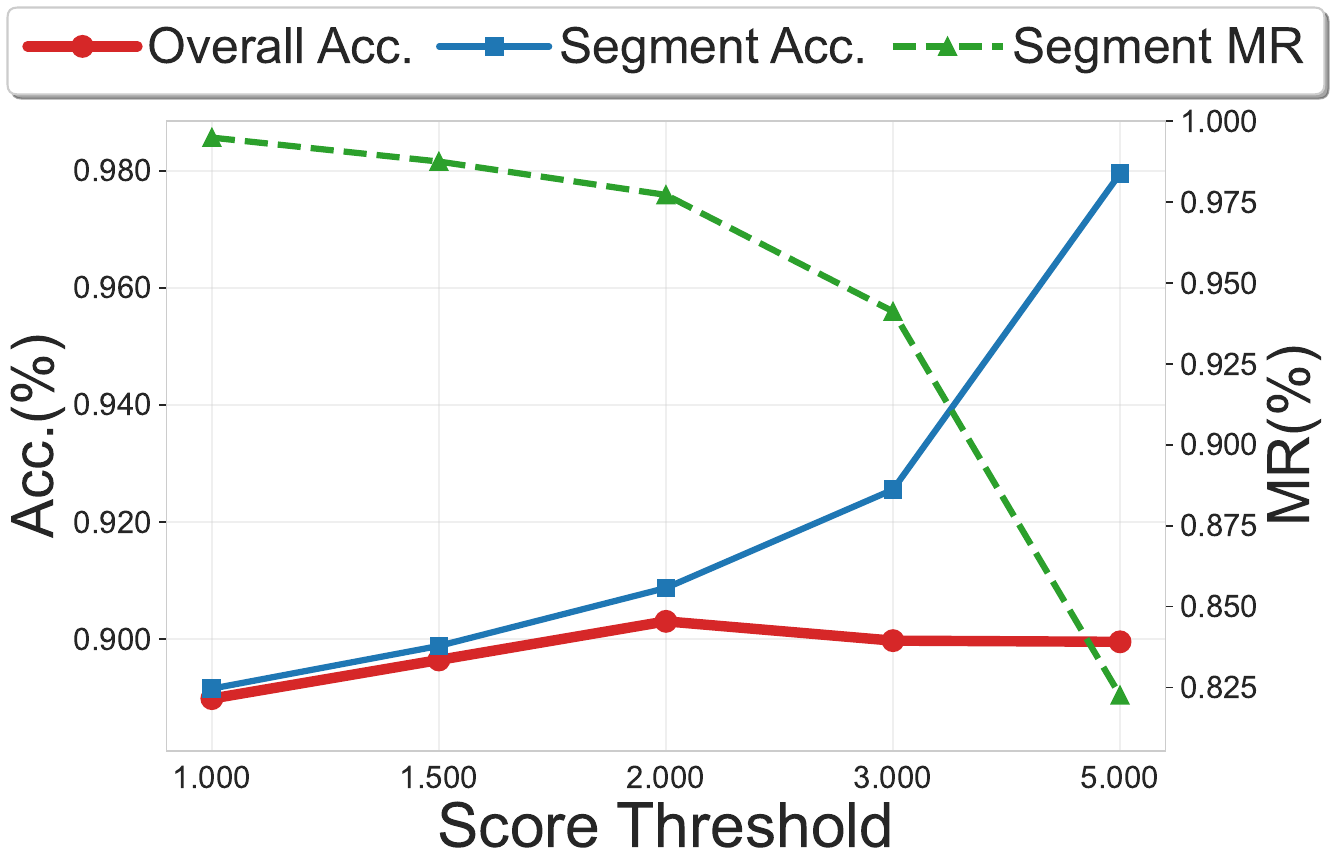}
    \caption{VisQUIC-10}
\end{subfigure}

\caption{Score threshold selection across datasets.}
\label{fig:threshold}
\end{figure}

Figure~\ref{fig:threshold} demonstrates the threshold selection trade-off across our evaluation datasets. As the score threshold increases, segment accuracy (Segment Acc.) consistently improves because only class-specific segments are retained, filtering out segments that might match samples from other categories. However, this improvement comes at the cost of reduced segment matching rate (Segment MR), meaning more samples must rely on the less accurate Backup Classification Mechanism for classification. Consequently, the overall accuracy (Overall Acc.) of Synecdoche reaches its optimal performance at specific threshold values: a score of 2.0 for three datasets (Bot-IoT, ToN-IoT, and CipherSpectrum-10) and 3.0 for VisQUIC-10. At these optimal thresholds, segment-based classification successfully handles 95\%, 85\%, 92\%, and 90\% of samples for Bot-IoT, ToN-IoT, CipherSpectrum-10, and VisQUIC-10, respectively.

\newcolumntype{C}[1]{>{\centering\arraybackslash}m{#1}}

\begin{table*}[t]
\centering
\caption{Classification Accuracy Comparison Across Different Traffic Types}
\label{tab:accuracy_comparison}
\renewcommand{\arraystretch}{1.2}
\begin{tabular}{l|C{1.0cm}C{1.0cm}|C{1.0cm}C{1.0cm}|C{1.0cm}C{1.0cm}|C{1.0cm}C{1.0cm}}
\hline
\multirow{2}{*}{\textbf{Method}} & \multicolumn{2}{c|}{\textbf{Bot-IoT}} & \multicolumn{2}{c|}{\textbf{ToN-IoT}} & \multicolumn{2}{c|}{\textbf{CipherSpectrum-10}} & \multicolumn{2}{c}{\textbf{VisQUIC-10}} \\
\cline{2-9}
 & \textbf{Acc} & \textbf{F1} & \textbf{Acc} & \textbf{F1} & \textbf{Acc} & \textbf{F1} & \textbf{Acc} & \textbf{F1} \\
\hline
\multicolumn{9}{l}{\textit{Data Plane Methods}} \\
\hline
\textbf{Synecdoche} & \textbf{0.997} & \textbf{0.959} & \textbf{0.852} & \textbf{0.793} & \textbf{0.919} & \textbf{0.915} & \textbf{0.903} & \textbf{0.897} \\
NetBeacon & 0.993 & 0.895 & 0.644 & 0.645 & 0.655 & 0.681 & 0.771 & 0.732 \\
Brain-on-Switch & 0.995 & 0.960 & 0.813 & 0.708 & 0.838 & 0.835 & 0.739 & 0.714 \\
\hline
\multicolumn{9}{l}{\textit{Offline Baselines}} \\
\hline
Random Forest & 0.989 & 0.919 & 0.837 & 0.824 & 0.822 & 0.826 & 0.775 & 0.737 \\
FS-Net & 0.997 & 0.941 & 0.892 & 0.863 & 0.997 & 0.997 & 0.956 & 0.956 \\
\hline
\end{tabular}
\end{table*}

\begin{figure}[!htb]
\centering
\begin{subfigure}[b]{0.22\textwidth}  
    \centering
    \includegraphics[width=\textwidth]{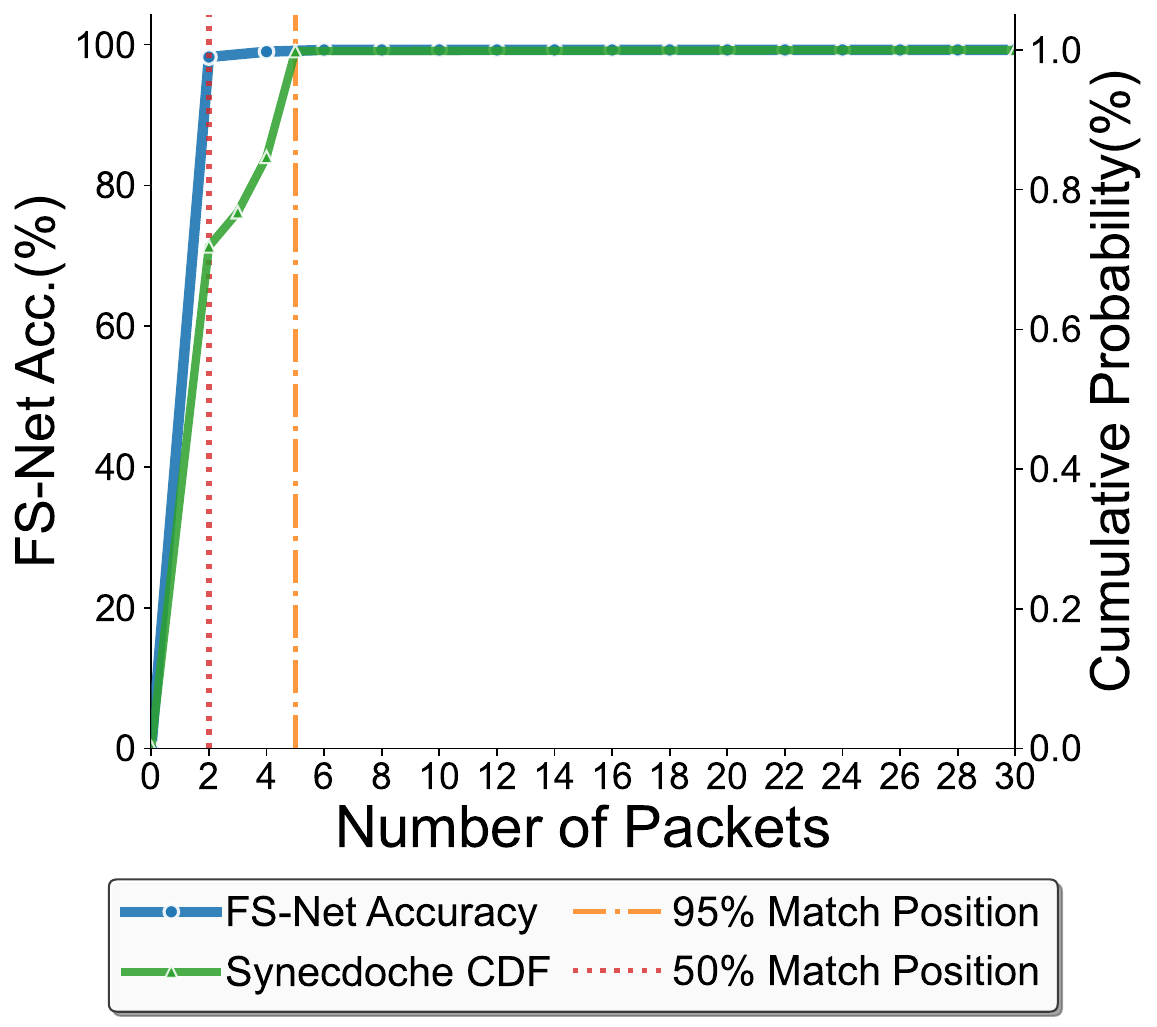}
    \caption{Bot-IoT}
\end{subfigure}
\quad  
\begin{subfigure}[b]{0.22\textwidth}
    \centering
    \includegraphics[width=\textwidth]{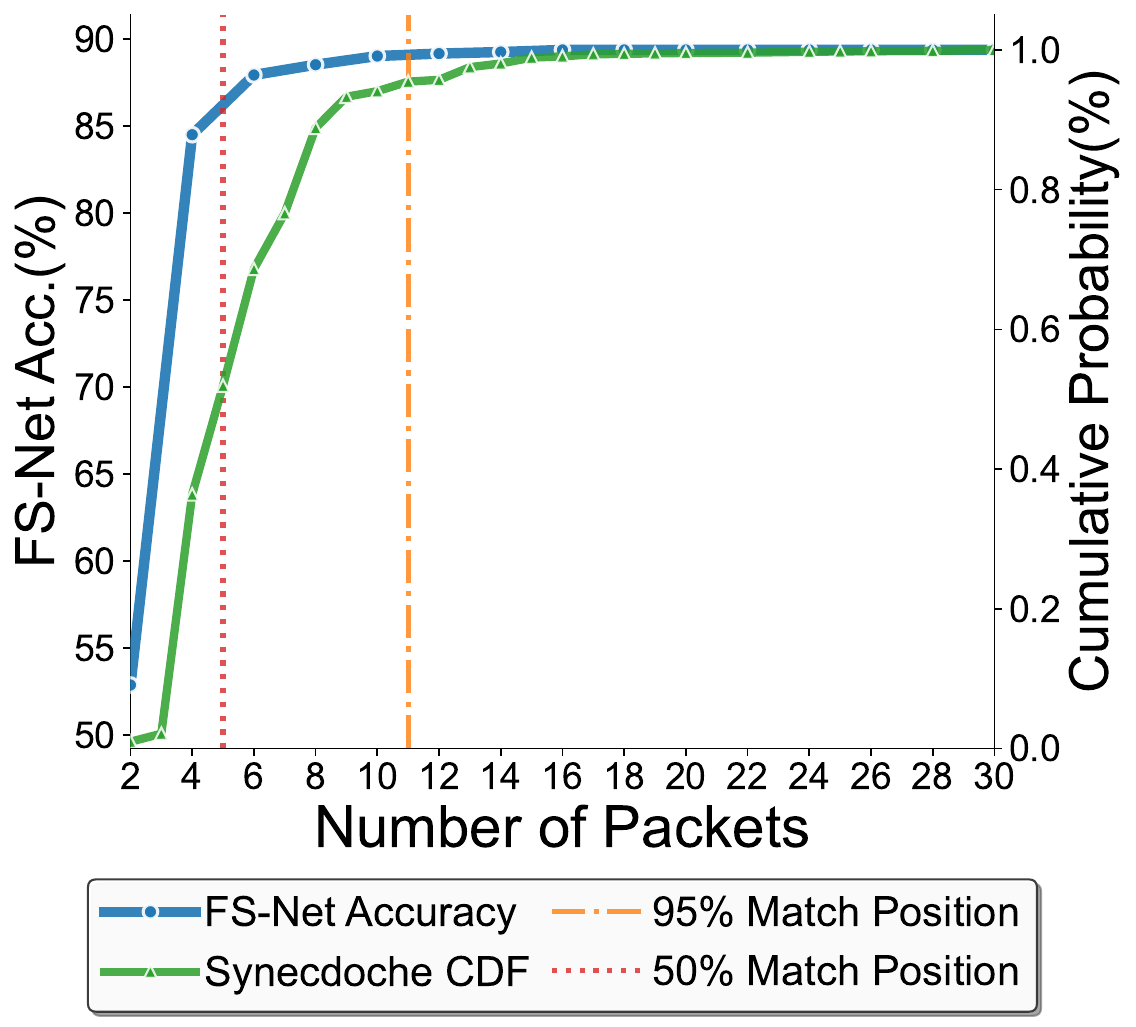}
    \caption{ToN-IoT}
\end{subfigure}

\medskip  

\begin{subfigure}[b]{0.22\textwidth}
    \centering
    \includegraphics[width=\textwidth]{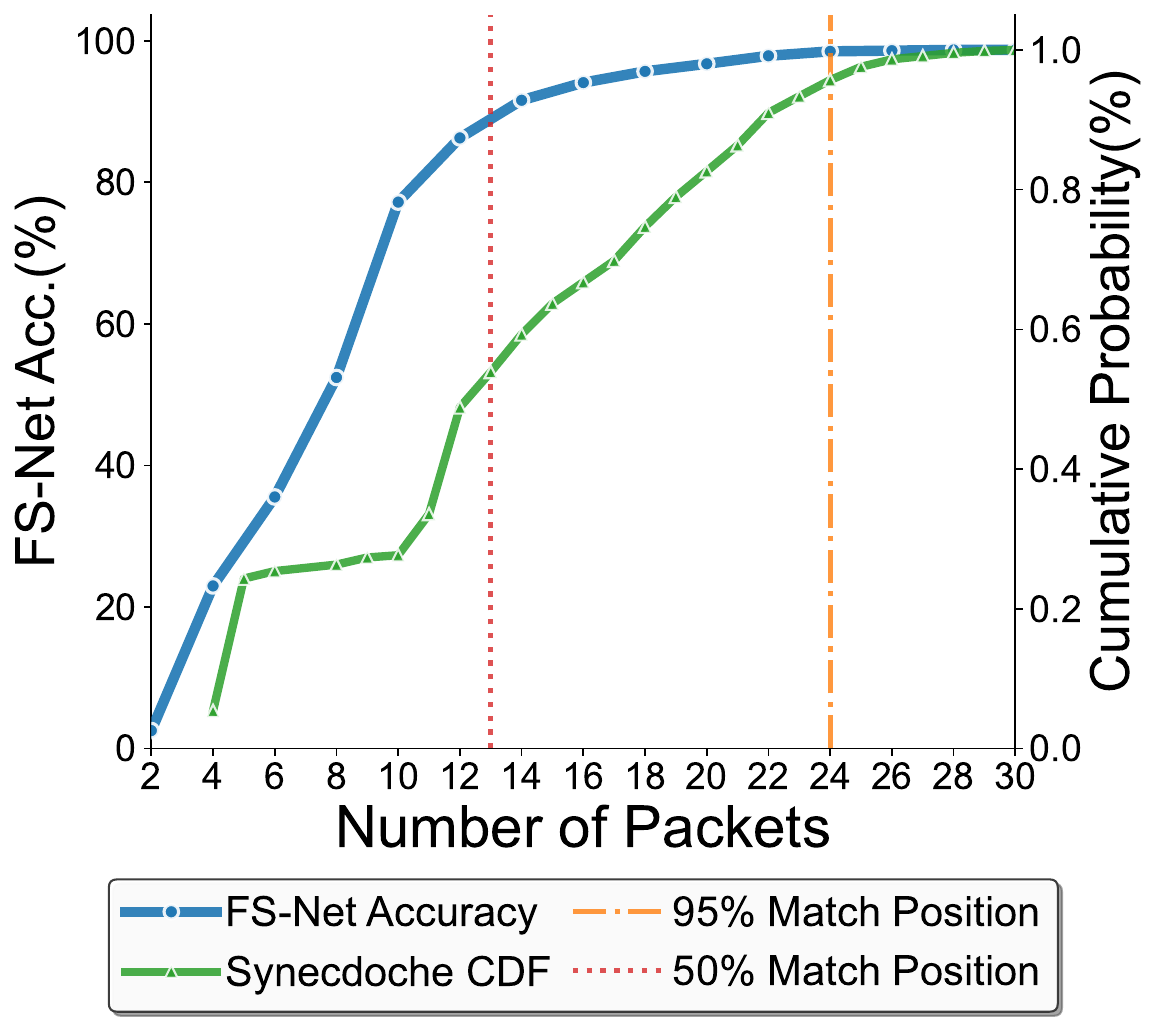}
    \caption{CipherSpectrum-10}
\end{subfigure}
\quad
\begin{subfigure}[b]{0.22\textwidth}
    \centering
    \includegraphics[width=\textwidth]{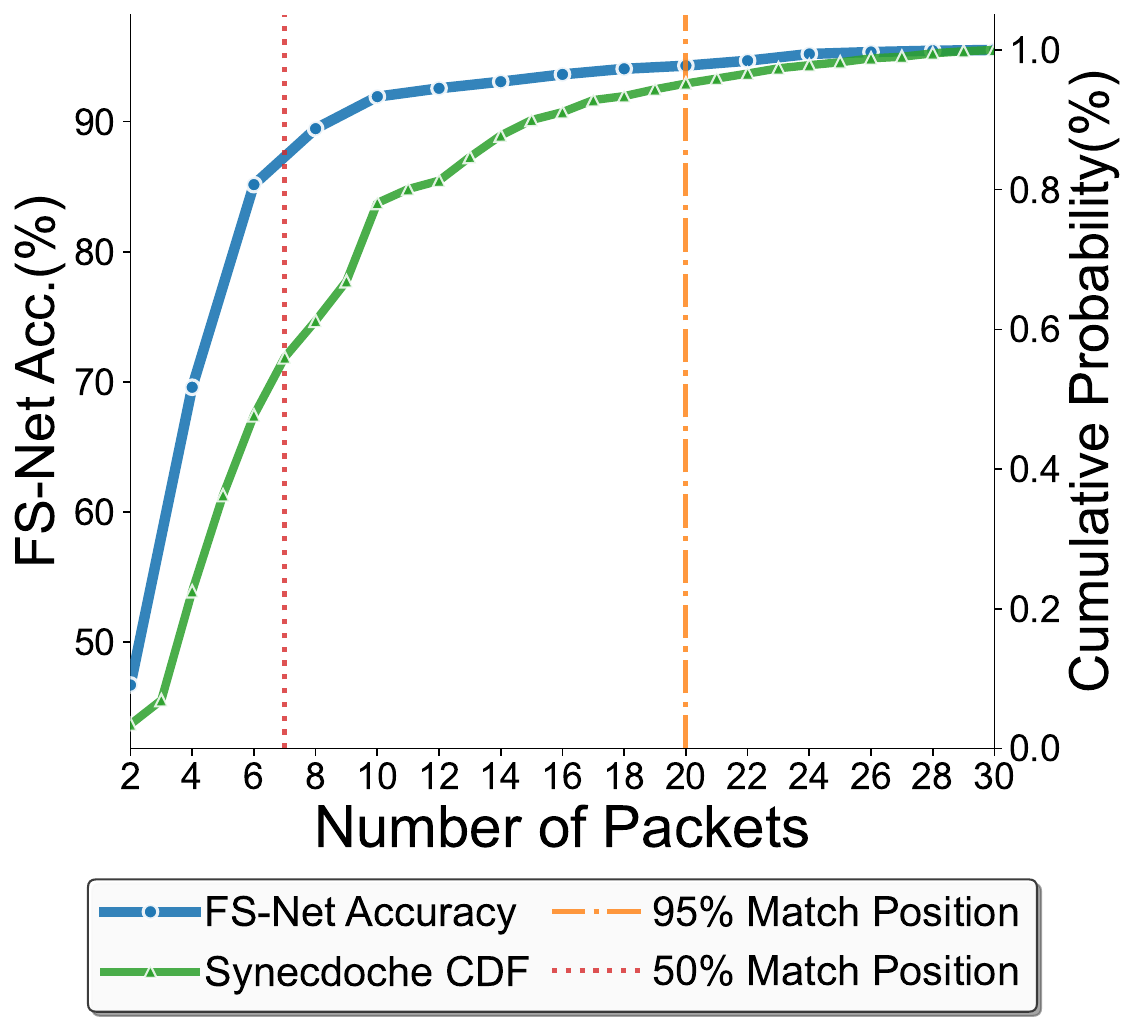}
    \caption{VisQUIC-10}
\end{subfigure}
\caption{Validation of Key Segment positioning through comparison with FS-Net accuracy convergence across different datasets.}
\label{fig:cdf}
\end{figure}

Figure~\ref{fig:cdf} demonstrates that Synecdoche's Key Segment matching positions closely align with FS-Net's accuracy convergence points. On ToN-IoT, 50\% of samples are classified by packet 6, where FS-Net begins stabilizing, while 95\% are classified by packet 12 near full convergence. This alignment confirms that Key Segments capture the same discriminative information used by deep learning models. Beyond validation, classification at these critical positions enables significant computational savings through early stopping while maintaining accuracy and reducing latency. Furthermore, different datasets show varying Key Segment positions: Bot-IoT (95\% by packet 10), ToN-IoT (packet 12), CipherSpectrum-10 (packet 22), and VisQUIC-10 (packet 20). This variation reflects the inherent characteristics of each dataset— attack detection datasets require fewer packets due to distinct malicious signatures, while application classification needs more packets to distinguish similar behaviors.

\subsection{Accuracy Comparison (RQ1)}

To answer RQ1, we evaluate the classification accuracy of Synecdoche compared to existing data plane methods and offline baselines across different traffic classification scenarios and class complexities.

Table~\ref{tab:accuracy_comparison} demonstrates that our method significantly outperforms existing data plane approaches across most datasets. Compared to the statistical feature-based NetBeacon, we observe substantial F1-score improvements: 22.9\% relative improvement on ToN-IoT (from 0.645 to 0.793) and 34.4\% on CipherSpectrum-10 (from 0.681 to 0.915). Against the sequential feature-based Brain-on-Switch (BoS), our method demonstrates superiority on three datasets, with F1-score improvements of 8.5\% on ToN-IoT, 8.0\% on CipherSpectrum-10, and 18.3\% on VisQUIC-10, while achieving comparable performance on Bot-IoT.

The results also validate that packet sequential features methods possess higher accuracy ceilings than statistical approaches. The offline comparison shows FS-Net consistently outperforming Random Forest, while among online methods, Brain-on-Switch generally surpasses NetBeacon, confirming that packet sequential features provide superior discriminative power for traffic classification tasks.

Furthermore, our method demonstrates superior performance in application classification scenarios for two main reasons. First, application-layer traffic exhibits more distinctive segment characteristics due to standardized communication patterns and fixed interface calls, making segment matching particularly effective. Second, as shown in Figure \ref{fig:cdf}, critical discriminative information for application classification typically resides in later packet positions within flows, while competing methods like NetBeacon (which makes decisions at packets 2, 4, 8, 16, etc.) and BoS (using multiple fixed sliding windows) often trigger premature classification decisions before sufficient contextual information becomes available. Our approach only initiates classification upon Key Segments detection, ensuring adequate information for accurate classification.

\begin{figure}[!htb]
\centering
\begin{subfigure}[b]{0.23\textwidth}  
    \centering
    \includegraphics[width=\textwidth]{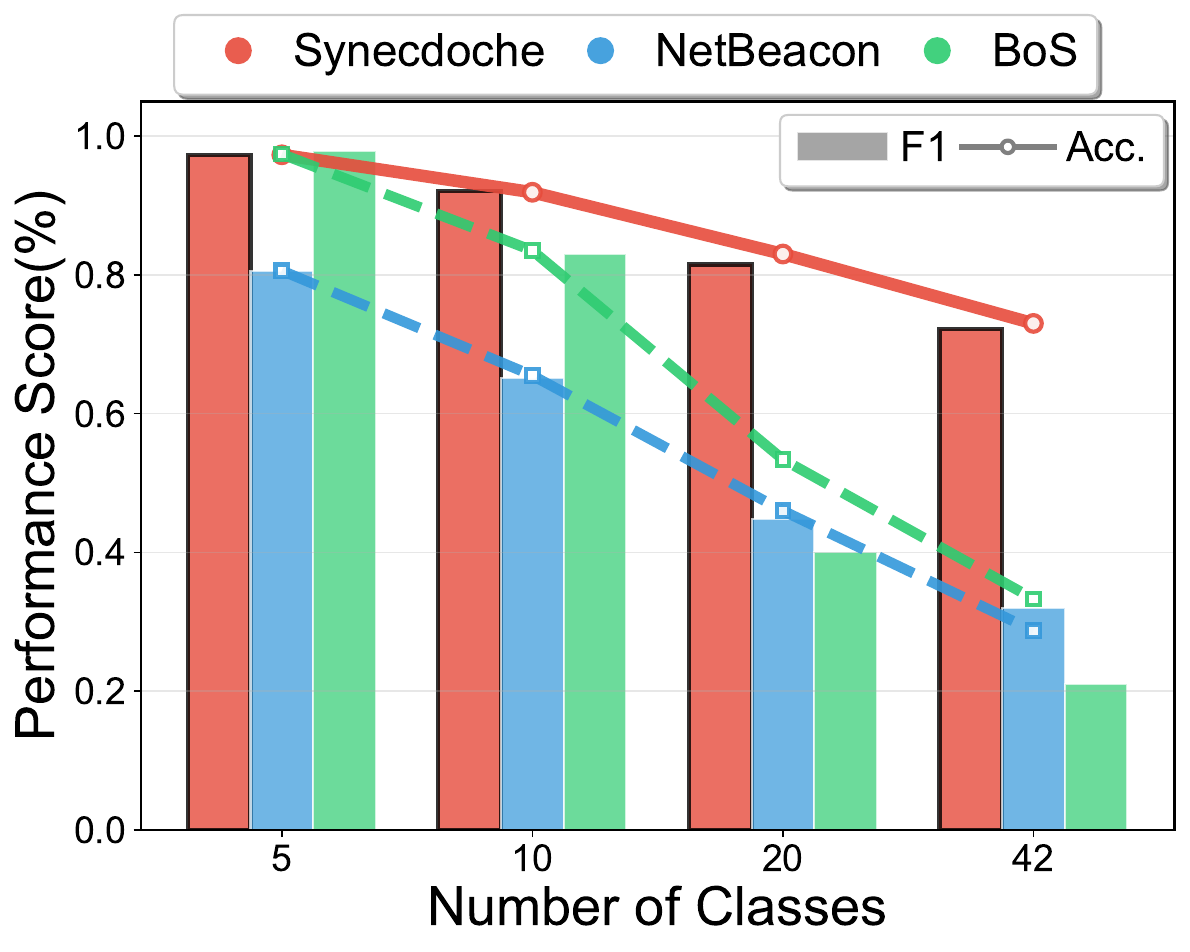}
    \caption{CipherSpectrum}
\end{subfigure}
\quad  
\begin{subfigure}[b]{0.23\textwidth}
    \centering
    \includegraphics[width=\textwidth]{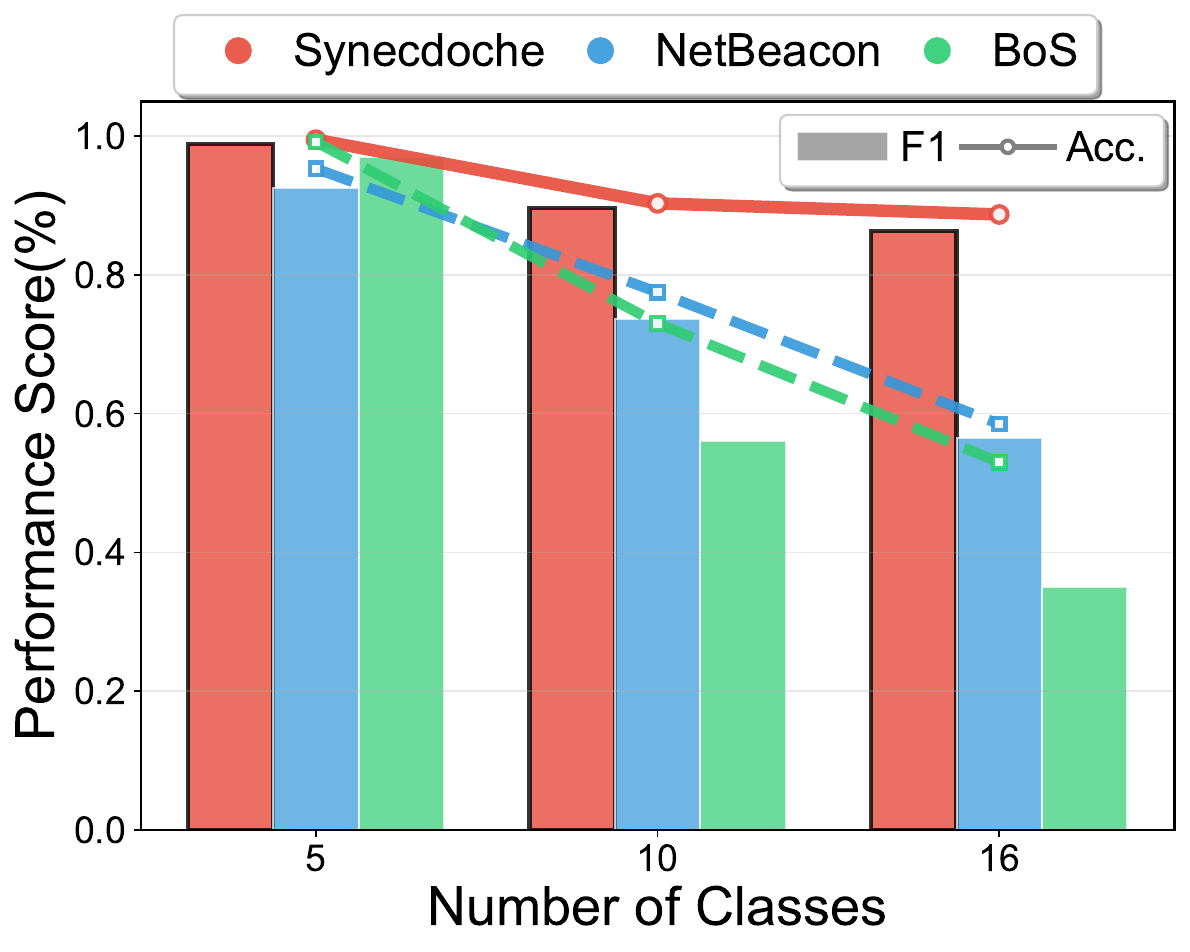}
    \caption{VisQUIC}
\end{subfigure}

\caption{Comparison of classification performance with increasing number of classes}
\label{fig:classes}
\end{figure}

To evaluate how our method's accuracy scales with increasing classification complexity, we conduct experiments across different numbers of classes for both application classification datasets. As demonstrated in \ref{fig:classes}, our Synecdoche method consistently outperforms both baseline approaches across all scenarios. On CipherSpectrum, our method maintains accuracy of 97.3\%, 91.9\%, 83.8\%, and 73.6\% for 5, 10, 20, and 42 classes, respectively. In contrast, NetBeacon degrades dramatically from 80.6\% to 28.7\%, and BoS drops from 97.4\% to 33.3\%. Similarly, on VisQUIC, our approach achieves 99.5\%, 90.3\%, and 88.7\% accuracy for 5, 10, and 16 classes, compared to NetBeacon's decline from 95.3\% to 58.5\% and BoS's deterioration from 99.2\% to 53\%. These results demonstrate the remarkable resilience of our method under increasing classification complexity, proving its strong potential for real-world multi-class application classification tasks.

\textbf{Answer to RQ1}: Synecdoche's direct matching of Key Segments not only achieves but surpasses sophisticated online deep learning methods like Brain-on-Switch, with F1-score improvements of 8.5\% on ToN-IoT and 18.3\% on VisQUIC-10, while significantly outperforming statistical approaches like NetBeacon. Additionally, Synecdoche demonstrates superior scalability under increasing classification complexity.

\begin{table}[h]
\centering
\caption{Rules Complexity among Datasets}
\renewcommand{\arraystretch}{1.2}
\label{tab:dataset_stats}
\begin{tabular}{lcc}
\hline
\textbf{Metric} & \textbf{UNSW\_ToN\_IoT} & \textbf{Cipher\_Spectrum} \\
\hline
Number of Classes & 10 & 42 \\
Range Match Rules (TCAM) & 359 & 1485 \\
Rules per Class & 35.90 & 35.53 \\
Exact Match Rules (SRAM) & 1102 & 7223 \\
Avg. Splits per Rule & 3.07 & 4.86 \\
\hline
\end{tabular}
\end{table}

\subsection{Performance and Efficiency Analysis (RQ2)}
To answer RQ2, we evaluate Synecdoche's efficiency in terms of latency and hardware resource utilization compared to existing data plane inference methods. Table~\ref{tab:dataset_stats} summarizes the rule statistics among datasets. With an average of 30+ rules per class, classification is achieved effectively. After splitting range matches into exact matches, no exponential explosion occurs—each range rule expands to 3-4 exact match rules on average. We allocate 2048 TCAM and 8192 SRAM entries accordingly, with all subsequent comparisons using this configuration. 

Table~\ref{tab:performance_comparison} presents comprehensive performance metrics. In terms of per-flow storage requirements, our approach maintains the feature information of the most recent $L_{max}$ packets, including: a 32-bit direction register storing the source IP of the first packet for subsequent direction determination, a 32-bit register for the latest packet timestamp, an 8-bit counter for the current packet count, and $L_{max} \times 16$ bits for packet features. When $L_{max}=4$, each flow requires 136 bits of storage, significantly outperforming both baselines. NetBeacon stores different statistical features totaling 272 bits per flow, while BoS requires 288 bits to maintain sliding window hidden states (window size=8) plus per-class cumulative scores for classification decisions. Both Synecdoche variants demonstrate superior resource efficiency in terms of SRAM utilization. The SRAM version achieves exceptional memory efficiency with only 4.03\% SRAM usage, representing a remarkable 79.2\% reduction compared to NetBeacon (19.38\%) and an 85.8\% reduction compared to BoS (28.33\%). 

Considering that switch hardware limits both the ingress and egress pipelines to 12 stages each, our Synecdoche variants demonstrate exceptional efficiency by completing the entire classification pipeline within a single pipeline direction. The SRAM version requires only 8 stages while the TCAM version uses 11 stages, both operating entirely within the ingress pipeline. NetBeacon also operates within the single-direction constraint using 12 stages. In contrast, BoS requires 24 total stages that must span both the ingress (12 stages) and egress (12 stages) pipelines. Both Synecdoche implementations achieve superior latency characteristics. The SRAM version delivers a per-packet processing time of 416 ns and the TCAM version 420 ns, which is 13.0\% faster than NetBeacon (470 ns) and 125.2\% faster than BoS (937 ns). This demonstrates the efficiency benefits of our segment-matching-based approach over complex statistical computation and deep learning inference.

\begin{table}[t]
\centering
\caption{Performance Comparison of Different Methods}
\label{tab:performance_comparison}
\renewcommand{\arraystretch}{1.2}
\setlength{\tabcolsep}{4pt} 
\begin{tabular}{l|c|c|c|c|c}
\hline
\textbf{Method} & \textbf{Bits/} & \textbf{SRAM} & \textbf{TCAM} & \textbf{Stages} & \textbf{Latency} \\
 &\textbf{Flow} & \textbf{(\%)} & \textbf{(\%)} & & \textbf{(ns)} \\
\hline
NetBeacon& 272 & 19.38 & 31.25  &12  & 470 \\
BoS & 288 &  28.33 & 6.94 &24 & 937 \\
Synecdoche (TCAM)  & 136 & 1.47 & 20.4 &11 & 420\\
Synecdoche (SRAM) & 136 & 4.03 & 1.09 &8  & 416 \\
\hline
\end{tabular}
\end{table}

\textbf{Answer to RQ2}: Synecdoche demonstrates superior performance in both latency and hardware resource utilization. It achieves 13.0\% faster processing than NetBeacon and 125.2\% faster than BoS, while reducing SRAM usage by 79.2\% compared to NetBeacon and 85.8\% compared to BoS, demonstrating exceptional memory efficiency.

\section{Conclusion}
This paper presents Synecdoche, a novel traffic classification framework that successfully bridges the accuracy-efficiency gap on programmable data plane through direct packet sequential pattern matching. By leveraging an "offline discovery, online matching" paradigm that automatically extracts Key Segments using deep learning models and deploys them as optimized table entries, Synecdoche significantly outperforms existing statistical and deep learning approaches while substantially reducing hardware resource consumption and processing latency. This work represents a fundamental advance in programmable data plane traffic classification, demonstrating that intelligent pattern extraction and hardware-optimized deployment strategies can overcome traditional accuracy-efficiency trade-offs and pave the way for sophisticated packet sequence analysis at line rate in emerging network applications. Our future work will continue to explore the impact of network fluctuations on Key Segment effectiveness and investigate mechanisms for efficient rule updates in deployed systems. The authors have provided public access to their code at https://github.com/swampx/Synecdoche.
\\
\section*{Acknowledgment}
This work was partly supported by The Cybersecurity Professional Committee of the Chinese Association of Educational Technology(NO. 2023CAET1002)

\vspace{12pt}

\color{red}

\end{document}